# Silicon rebirth: Metallic prismane allotropes of silicon for the next-gen technologies

Konstantin P. Katin[1,2], Konstantin S. Grishakov[1], Margarita A. Gimaldinova[1], and Mikhail M. Maslov[1,2,*]

*e-mail: Mike.Maslov@gmail.com

[1]Nanoengineering in Electronics, Spintronics and Photonics Institute, National Research Nuclear University "MEPhI", Kashirskoe Shosse 31, Moscow 115409, Russia

[2]Laboratory of Computational Design of Nanostructures, Nanodevices, and Nanotechnologies, Research Institute for the Development of Scientific and Educational Potential of Youth, Aviatorov str. 14/55, Moscow 119620, Russia

**Abstract**

We report the prediction of metallic quasione-dimensional $sp^3$-hybridized silicon allotropes in the form of prismanes. Silicon prismanes or polysilaprismanes are the silicon nanotubes of a special type constructed from the dehydrogenated molecules of cyclosilanes (silicon rings). By means of density functional theory the electronic, geometry, energy, and some mechanical properties of these tubes were investigated. Our results show that silicon polyprismanes are thermodynamically stable compounds, and the character of the energy spectrum, as well as the behavior of transmission function near the Fermi level, show that they exhibit non-typical for the silicon systems metallic nature. Moreover, the metallic state of polysilaprismanes is resistant to the mechanical stresses applied along their main axis. Unusual properties predicted in the presented study discover new prospects of application of silicon nanostructures as the basic elements of future micro- and nanoelectronics, as well as in energy, metrology, medical, and information technologies.

**Introduction**

For a long time to the present days, diamond-type silicon is the basis of modern consumer and special purpose electronics. However, the active development of nanotechnologies makes it possible to think about the use of silicon in new forms, for example, in the form of polyprismanes, which can qualitatively improve the component base of modern electronic devices. Silicon polyprismanes or [$n,m$]silaprismanes can be considered as the layered dehydrogenated molecules of cyclosilanes (silicon rings), where $m$ is the number of vertices of a closed silicon ring, and $n$ is the number of layers. For a large $n$, polysilaprismanes can be considered as the analogs of silicon single-walled nanotubes with an extremely small cross-section in the form of a regular polygon.

At the moment, silicon polyprismanes, unlike their carbon analogs, have not been studied in sufficient detail, although the other quasione-dimensional silicon tubular nanostructures have been successfully analyzed [1]. Moreover, the idea of replacing carbon atoms in traditional prismanes with atoms of other elements or even functional groups has already been expressed in the scientific literature [2]. It is found that the substitution of carbon atoms by silicon atoms stabilizes the framework of prismane cage [3], which is further confirmed by computer simulations of the comparative analysis of the kinetic stability of small carbon and silicon prismanes [4]. Unlike carbon prismanes, even small silicon prismanes are able to contain heavy atoms in a silicon cell, for example, silicon or germanium, maintaining the structure of the prismane, which will definitely have a positive effect on the possible formation of the endohedral compounds based on host-guest type [4,5] that will be useful in biomedicine and photovoltaics [6]. Despite the fact that the simplest representatives of polysilaprismanes were synthesized long ago [7-9], unfortunately, to date, there has not been developed a universal method for the obtaining these molecular systems, therefore, higher polysilaprismanes are still being investigated exclusively theoretically, including the computer simulation. Thus, the thermodynamic stability of these systems, including small endohedral silaprismanes [10-13], the structural characteristics of higher silaprismanes are calculated and the dependence of their some electronic properties on the effective length of polysilaprismanes is confirmed [14], the influence of an embedded atom on the optical properties of small prismanes is established, and the questions of their kinetic stability are studied [4]. It has been found that the doping with certain atoms (for example, fluorine atoms) is capable of altering the mechanical properties of polysilaprismanes [15]. However, despite a number of achievements in the physicochemistry of polysilaprismanes, many questions still remain open. For example, an analysis of the thermal stability and chemical reactivity of polysilaprismanes, the study of channels and decay products, could suggest possible ways of their synthesis. Obtaining new knowledge about the electronic characteristics of polysilaprismanes would allow one to make the scientifically based conclusions about the prospects of using them in next-gen technologies instead of traditional silicon materials.

To the best of our knowledge, it is the first study of quasione-dimensional $sp^3$-hybridized silicon complexes. We have calculated thermodynamic stability, electronic properties, and chemical reactivity of silicon [*n,m*]prismanes with $m = 5$, 6, 7, and 8 containing from two to ten layers by means of calculating frontier molecular orbitals, and some reactivity descriptors. Furthermore, we have performed structural optimization of rather long ("infinite") polyprismanes taking into account periodic boundary conditions and obtained their band structure, density of electronic states, electronic transmission coefficients, and some mechanical properties to predict their behavior as the part of perspective nanotechnology devices and to illustrate the potential of these systems for the future development of next-gen nanotechnologies. Thus, this study is the first step toward analyzing the physicochemical properties of silicon tubular nanostructures based on the extended polysilaprismane molecules.

## 1. Materials and Methods

### 1.1. Molecular Structure of Silicon Polyprismanes

We have analyzed a family of finite silicon [$n,m$]prismanes with $m$ = 5, 6, 7, and 8 up to ten layers ($n$ = 2 ÷ 10) as well as rather long ("infinite") polysilaprismanes under periodic boundary conditions. Molecular structures of finite systems are displayed in Figure 1. Hydrogen passivation at the ends of polysilaprismanes was used to avoid dangling bonds.

For the polysilaprismanes modeling under the periodic boundary conditions we used a corresponding supercell containing four elementary cells, i.e., four interconnected silicon regular polygons ($Si_5$ for the [$n$,5]-, $Si_6$ for the [$n$,6]-, $Si_7$ for the [$n$,7]-, and $Si_8$ for the [$n$,8]polysilaprismane).

### 1.2. Computational Details

Precise optimizations and characterization of silicon polyprismanes were made in terms of density functional theory (DFT). All DFT calculations for the finite [$n,m$]silaprismanes were performed using the TeraChem software [16-19]. The PBE0 functional (hybrid analog of traditional Perdew, Burke and Ernzerhof GGA PBE-functional [20]) and traditional Becke's three-parameter hybrid method and the Lee-Yang-Parr exchange-correlation energy functional (B3LYP) [21,22] were used. PBE0 functional was combined with Pople basis set 6-311G(d,p) [23], and B3LYP functional was combined with Dunning's correlation-consistent basis set aug-cc-pVDZ [24,25], as implemented in TeraChem. During geometry optimization, the global charge of all systems considered was neutral. The maximum force and root mean square forces were $4.5 \times 10^{-4}$ and $3 \times 10^{-4}$ (hartrees/bohr and hartrees/radian), whereas the maximum displacement and root mean square displacement were $1.8 \times 10^{-3}$ and $1.2 \times 10^{-3}$. All electronic characteristics and reactivity descriptors were calculated at the same PBE0/6-311G(d,p) and B3LYP/aug-cc-pVDZ levels of theory.

The DFT calculations of polysilaprismanes under periodic boundary conditions were carried out using the generalized gradient approximation (GGA) [20], with the PBE functional for the exchange-correlation terms and ultrasoft pseudopotential for silicon [26] with a cutoff energy value of 60 Ry. These calculations were performed using the Quantum Espresso v. 6.3 program package [27]. For the silicon polyprismane representation, we used the supercell containing four silicon regular polygons (pentagons, hexagons, heptagons, or octagons) periodically repeated along the z-axis (along the main axis of polyprismane). A Monkhorst-Pack [28] $k$-point grid of 1×1×20 and the first-order Methfessel-Paxton scheme [29] with a smearing of 0.01 Ry were implemented. Note that the choice of higher values of the kinetic energy cutoffs or denser $k$-point grids leads to the minor changes in the structural parameters, total energy, and electronic properties of the considered nanostructures. For example, the use of the cutoff energy value of 100 Ry and 1×1×40 $k$-point grid leads to the changes in total energy not exceeding the value of $7 \times 10^{-4}$ Ry. The structural optimization was performed until all components of the forces acting on the atoms became smaller than the value of $10^{-4}$ eV/Å. Note that the supercell parameters in z-direction were also optimized, while in the nonperiodic XY-plane a large vacuum space (more than 14 Å in every direction) was introduced between the Si-atoms to eliminate the nonphysical interactions.

1.3. Chemical Reactivity Descriptors

Density functional theory was used in this study for understanding of electronic properties and chemical reactivity of finite silicon polyprismanes up to ten layers, and for obtaining a set of physical quantities such as the electron affinity (EA) [30,31], first ionization potential (IP) [30,31], chemical hardness ($\eta$) [32] and softness ($S$) [31], chemical potential ($\mu$) [33], electronegativity ($\chi$) [34,35], and electrophilicity index ($\omega$) [33], which are linked to their electronic structure. Chemical potential, hardness, and electronegativity are defined as follows

$$\mu = \left(\frac{\partial E}{\partial N}\right)_v, \qquad (1)$$

$$\eta = \frac{1}{2}\left(\frac{\partial^2 E}{\partial N^2}\right)_v = \frac{1}{2}\left(\frac{\partial \mu}{\partial N}\right)_v, \tag{2}$$

$$\chi = -\mu = -\left(\frac{\partial E}{\partial N}\right)_v, \tag{3}$$

where $E$ and $v$ are the electronic energy and the constant external potential of an $N$-electron system. As one can see from Eq. (2) chemical hardness is the resistance of the chemical potential to change in the number of electrons, i.e., it describes the resistance of the system to exchange electronic charge with the environment [31]. The concept of electronegativity was introduced by Pauling as the power of an atom in a molecule to attract electron to it. The Koopmans' theorem for closed-shell molecular systems [36] was used for the calculation of these parameters. According to this theorem, the ionization potential and electron affinity can be defined as the negative of the highest occupied molecular orbital (HOMO) $\varepsilon_H$ and lowest unoccupied molecular orbital (LUMO) $\varepsilon_L$ energy, respectively. Electron affinity refers to the capacity of the molecular systems to accept one electron from the donor while the ionization potential refers to the ability of the molecular systems to lose electrons. Thus, the use of the finite difference approximation and Koopmans' theorem leads to the following expression for the chemical potential

$$\mu = -\frac{1}{2}(\text{IP} + \text{EA}) = \frac{1}{2}(\varepsilon_L + \varepsilon_H). \tag{4}$$

In addition, hardness can be calculated as

$$\eta = \frac{1}{2}(\varepsilon_L - \varepsilon_H) = \frac{1}{2}\Delta_{HL}, \tag{5}$$

where $\Delta_{HL} = \varepsilon_L - \varepsilon_H$ is the HOMO-LUMO gap that is defined as the energy gap between the highest occupied molecular orbital and the lowest unoccupied molecular orbital. Note that the maximum electronic charge that the system can accept is $\Delta N_{max} = -\mu/\eta$ [33]. Softness is logically the reciprocal of hardness [31]

$$S = \frac{1}{2\eta}. \tag{6}$$

The electrophilicity concept was proposed by Parr and co-workers, and electrophilicity index is defined as [33]

$$\omega = \frac{\mu^2}{2\eta}. \tag{7}$$

Electrophilicity index measures the stabilization in energy when the system acquires an additional electronic charge from the environment. According to Parr, the analog can be drawn between Eq. (7) and the equation for power in classical electricity. So, the electrophilicity index can be described as some kind of "electrophilic power". Since electronegativity is additive inverse of the chemical potential, the electrophilicity index can be also written as

$$\omega = \frac{\chi^2}{2\eta}. \tag{8}$$

We obtained all above mentioned quantum-chemical descriptors for the finite silicon [$n,m$]prismanes with $m$ = 5, 6, 7, 8, and up to ten layers ($n$ = 2 ÷ 10) in a length.

## 2. Results and discussion

### 2.1. Geometric and energy characteristics

In the beginning, we construct a family of silicon [$n,m$]prismanes with $m$ = 5, 6, 7, 8 and $n$ = 2 ÷ 10 (see Figure 1). So, for the finite nanostructures, we restrict ourselves to the maximum of ten layers in polyprismane. Note that we do not take into consideration the [$n,4$]polysilaprismanes in this study, because as it was shown earlier the rather long prismanes with $n$ = 10 or higher were not thermodynamically stable [37].

First, we optimize and analyze the geometry of quasione-dimensional silicon polyprismanes. To study the dependence of the structural characteristics on their effective sizes we calculate the interlayer ($l_{\parallel}$) and intralayer ($l_{\perp}$) Si–Si bonds (see Figure 1) as well as the average effective Si–Si bond distance between the layers for every prismane

$$\langle l_\| \rangle = \frac{\sum_{i=1}^{n-1}(l_\|)_i}{n-1}, \tag{9}$$

where *n* is the standard designation of the number of layers in the polyprismane. The obtained average effective Si–Si bond distances for silicon polyprismanes consisting of two to ten layers are presented in Figure 2. Interlayer Si-Si bonds of all polyprismanes one can find in the Supplementary Materials. From Figure 2 one can see that the increase of the polyprismanes length does not affect strongly their geometric characteristics. So, it is very likely that the structural properties of the "long" ($n \to \infty$) polyprismanes will remain the same. Geometric data obtained for the silicon polyprismanes under the periodic boundary conditions are presented in Table 1. One can see that data obtained for the finite systems are consistent with the data calculated under periodic boundary conditions.

**Table 1.** Interlayer ($l_\|$) and intralayer ($l_\perp$) Si–Si bonds for silicon polyprismanes and their effective diameters (*D*) obtained under periodic boundary conditions ($n \to \infty$).

| *m* | 5 | 6 | 7 | 8 |
|---|---|---|---|---|
| $l_\|$, Å | 2.421 | 2.420 | 2.428 | 2.423 |
| $l_\perp$, Å | 2.402 | 2.393 | 2.389 | 2.388 |
| *D*, Å | 4.087 | 4.786 | 5.505 | 6.239 |

Next, we calculated the binding energies $E_b$ of the silicon polyprismanes. The binding energy $E_b$ of the polyprismane per atom is determined by the equation

$$E_b\left[\frac{\text{eV}}{\text{atom}}\right] = \frac{1}{N_{at}}\{kE(\text{H}) + lE(\text{Si}) - E_{tot}(\text{SPP})\}, \tag{10}$$

where $N_{at} = k + l$ is the total number of atoms in the polyprismane, $E_{tot}(\text{SPP})$ is the total silicon polyprismane energy, $E(\text{H})$ and $E(\text{Si})$ are the energies of the isolated hydrogen and silicon atoms, respectively. The polyprismane with higher binding energy (lower potential energy) is more thermodynamically stable and vice versa. The binding energies $E_b$ obtained for silicon polyprismanes consisting of two to ten layers are presented in Figure 3. The values of binding energies of all silicon polyprismanes considered one can find in the Supplementary Materials. As evident from Figure 3, as the number of silicon layers in the polyprismane increases, binding energy becomes larger. On the other hand, if the number of vertices of regular polygons *m* that make up the polyprismane increases, the binding energy slightly reduced (3D plot of the binding energy dependence on *n* and *m* one can find in Supplementary Materials). In general, the increase of binding energy indicates that silicon polyprismanes become more thermodynamically stable as their effective lengths increase. Maybe the high thermodynamic stability of "long" polyprismanes is associated with the decreasing of boundary induced structural distortions as the number of Si-rings increases. Thus, the formation of macroscopic 1D-nanoneedles in the form of silicon polyprismanes may be energetically favorable. However, the question of their kinetic stability requires an additional study.

## 2.2. Electronic properties

To understand the electronic properties of silicon polyprismanes, first, we calculate the HOMO-LUMO gaps of finite silicon [*n*,*m*]prismanes with *m* = 5, 6, 7, 8 and *n* = 2 ÷ 10. The HOMO-LUMO gap $\Delta_{HL}$ is defined as the energy gap between the highest occupied molecular orbital and the lowest unoccupied molecular orbital. The obtained results for HOMO-LUMO gaps are presented in Figure 4. HOMO and LUMO energies themselves of all silicon polyprismanes considered one can find in Supplementary Materials. The HOMO-LUMO gap analysis is very important for the proposed use of a material in electronic applications as well as for the estimation of its reactivity. As it is evident from Figure 4, the value of the HOMO-LUMO gap goes down with the increasing of the number of polyprismane layers *n*. A similar trend is predominantly maintained with increasing the number of vertices of a closed silicon rings *m* that make up the polyprismane. Note that the character of the HOMO-LUMO gap dependence on the number of layers in polysilaprismane remains the same when using a different level of theory. Figure 5 shows these dependencies for *m* = 5 and 6 at PBE0/6-311G(d,p) and B3LYP/aug-cc-pVDZ levels of theory. Such behavior of the HOMO-LUMO gap is typical for various quasione-dimensional nanostructures, for example, carbon [38] and boron nitride [39] nanotubes or traditional carbon polyprismanes [40]. In the limit of the bulk system, HOMO-LUMO gap aims to the energy gap $E_g$. In turn, the energy gap for bulk materials is a prime factor in determining its electrical conductivity σ. The relationship between σ and $E_g$ can be given by the following formula [41]

$$\sigma \propto \exp\left(-\frac{E_g}{k_B T}\right), \qquad (11)$$

where $k_B$ is the Boltzmann constant and $T$ is the temperature. One can see that a small decrease in $E_g$ leads to a significant increase in electrical conductivity. According to our evaluation of $\Delta_{HL}$, silicon polyprismanes can be characterized as the narrow-gap semiconductors/semimetals or even the metallic systems considering the significant decrease of HOMO-LUMO gap with the growth of the polyprismanes length. These $\Delta_{HL}$ values are small enough to allow electronic conductivity. Thus, in the bulk limit, it is possible to transport electrons, without any additional doping with donor/acceptor atoms or mechanical stresses.

To understand a metallicity of silicon polyprismanes in detail, we analyze the band structure, density of electronic states and electronic transmission coefficients of these nanostructures under periodic boundary conditions. The band structures, as well as the densities of the electronic states (DOS) for all systems considered, are presented in Figure 6. One can see that the energy gap characteristic for the semiconductors is absent for all considered polysilaprismanes. On the contrary, the traditional for metals overlap of the energy bands containing the Fermi level is observed. In the DOS plots, one can find the Van Hove singularities [42] that are characteristic for the quasione-dimensional nanostructures. Note that recently the prediction of a Na-Si clathrate structure have been reported, which can be used as a precursor for obtaining the pure metal bulk silicon phase at ambient pressure [43]. However, the family of silicon polyprismanes is a unique class of 1D metallic silicon compounds. Moreover, we analyze how the mechanical stretching and compression affects the electronic properties of gapless polysilaprismanes. The polysilaprismanes were stretched and compressed along their main axes by ten percent. It was found that such mechanical deformations do not open the energy gap. So, the metallicity of polysilaprismanes is resistant to mechanical deformations.

To confirm the metallic behavior of polysilaprismanes we calculate the transmission coefficients of electron tunneling through the polysilaprismanes using the PWcond module of Quantum Espresso. The PWcond code implements the scattering approach [44] for the study of coherent electron transport in atomic-sized "contacts" using the Landauer-Büttiker formalism [45-48]. The calculation of transmission coefficients was made without the bias voltage. The polysilaprismane was partitioned into a "device" consisting of supercell containing four elementary cells and the left and right "contacts" that were presented by the same supercells as well. According to our calculations, the number of ballistic channels increases with $m$ growth (see Figure 7). For example, $Si_5$- and $Si_6$-polyprismanes demonstrate six ballistic channels while $Si_7$- and $Si_8$-polyprismanes demonstrate ten ballistic channels, and therefore they possess an amount conductance of $6G_0$ and $10G_0$, respectively, where $G_0 = 2e^2/h$ is the conductance quantum (here $e$ is the elementary charge and $h$ is the Plank's constant). It should be noted that the electronic transmission around the Fermi energy of the common armchair (metallic) carbon nanotubes is sufficiently lower [49,50], the semiconductor carbon peapods become metals only under the influence of strong mechanical deformations [51], and the carbon analogs of silicon polyprismanes are wide-gap semiconductors [40].

### 2.3. Elastic properties

As noted above, mechanical deformations do not affect strongly the electronic properties of polysilaprismanes. However, we analyze the mechanical properties of silicon silaprismanes to investigate their response to both tensile and compressive strains along their main axes. We restrict ourselves to small strain values (not exceeding 4%) in order to maintain the system in the elastic region. For all considered polysilaprismanes we estimate Young's modulus that is given as the second derivative of the total energy $E$ with respect to the strain $\varepsilon$ divided by the equilibrium volume $V_0$ [52]

$$Y = \frac{1}{V_0} \frac{\partial^2 E}{\partial \varepsilon^2}. \qquad (12)$$

In the case of polysilaprismanes, we define the equilibrium volume as the volume of the hollow regular prism $V_0 = S_m z_0$, where $S_m$ is the area of the corresponding regular polygon ($m$ = 5, 6, 7, and 8) taking into account the covalent radius of the silicon atom $R_{Si} = 1.11$ Å, and $z_0$ is the equilibrium polysilaprismane length. Using the Eq. (12) we obtain Young's modules 192.8; 149.6; 124.7; 110.7 GPa for the $Si_5$-, $Si_6$-, $Si_7$-, and $Si_8$-polyprismanes, respectively. These values are rather high and comparable to the values obtained for some metals and steels as well as for the high-stable silicon tubular nanostructures [1]. Note that if the number of vertices of regular silicon polygons $m$ that make up the polyprismane increases the value of Young's modulus reduces. Thus, polysilaprismanes with smaller effective diameters are more resistant to the uniaxial compression and stretching under elastic deformation than the polysilaprismanes with larger effective diameters.

### 2.4. Chemical reactivity

The chemical reactivity indices (descriptors) obtained for the silicon polyprismanes are electron affinity, ionization potential, chemical hardness and softness, chemical potential, electronegativity, and electrophilicity index. Chemical potential, hardness, softness, and electrophilicity index of the silicon [$n,m$]prismanes with $m$ = 5, 6, 7, and 8 up to ten layers ($n$ = 2 ÷ 10) are presented in Figures 8-11, respectively. More complete data one can find in Supplementary Materials.

The chemical reactivity parameters are very important indices because they illustrate the reactivity and stability of a molecular complex. Electronic chemical potential characterizes the energy change when electrons are added to the molecular system, in other words, when the electron density changes. As one can see from Figure 8, the chemical potential dependencies on the number of silicon rings in the polysilaprismane have a "sawtooth" look. It can be said that the local maxima of chemical potential are observed when the number of silicon layers is a multiple of three for every $m$ (of course, if $n = 2$ is excluded from consideration). Since the electronegativity is defined as the negative of the electronic chemical potential, it has local maxima at $n = 4, 7$, and $10$ (see Figure 9). Electronegativity is an indicator of how favorable the molecular system is to the penetration of an electron charge. It is difficult to highlight the clear trend, but, in general, it can be said that the silicon polyprismanes with lower $m$ are more electronegative than the polysilaprismanes with higher $m$. In contrast, it is clear that the chemical hardness decreases with an increase in the number of layers $n$ in the polysilaprismane for every $m$ (see Figure 10), and the chemical softness (as the reciprocal value of hardness) increases. So, when increasing the number of layers the polysilaprismane is less and less hard. Chemical hardness characterizes the change rate of the electronic chemical potential with an electron density change. A hard molecular system is a system for which a small change in electron density leads to a significant change in the chemical potential. Therefore, it can be said that the ability of rather long polysilaprismanes to accept still more electrons is not compromised. In this regard, the silicon polyprismanes are prone to electronic configuration changes. The analysis of electrophilicity reveals that, in general, it increases with an increase in the number of polysilaprismane layers (Figure 11), and furthermore $Si_7$- and $Si_8$-polyprismanes are more electrophilic than the $Si_5$- and $Si_6$-polyprismanes at large $n$. On the other hand, the interesting behavior of electrophilicity index is observed at the number of silicon rings in the polyprismane $n = 9$ and $10$. From the Figure 11 one can see that [9,5]- and [9,6]polysilaprismanes are less electrophilic than [10,5]- and [10,6]polysilaprismanes, and, in contrary, [9,7]- and [9,8]polysilaprismanes are more electrophilic than [10,7]- and [10,8]polysilaprismanes. Moreover, at $n = 10$, the values of electrophilicity of polysilaprismanes with $m = 5, 6, 7$, and $8$ are quite close. So, it is difficult to distinguish the clear trend of electrophilicity behavior at the bulk limit ($n \to \infty$). However, it can be assumed that at "infinite" length the silicon polyprismanes would be quite electrophilic. In addition, it should be noted that the behavior of Parr electrophilicity index indicates that the stabilization in energy when the system acquires an additional electronic charge from the environment goes down with the increase of the effective sizes of the polysilaprismanes. Note that the maximum electronic charge $\Delta N_{max}$ that the polyprismane can accept also strongly depends on its structure (data are presented in Supplementary Materials). In general, $\Delta N_{max}$ for the silicon polyprismanes that have larger diameter is greater than for the thinner ones.

2.5. Application prospectives

The relevance of the proposed study is determined by the possibility of creating the fundamentally new functional nanomaterials that will form the basis of future micro- and nanoelectronics, energy, metrology, and information technologies. The extraordinary electronic properties of silicon polyprismanes together with their structural and mechanical characteristics make them potentially suitable in the various fields of next-gen technologies. Building the next generation of technological processes is impossible without the use of an improved material base. For example, the miniaturization of electronics inevitably leads to the using of appropriate semiconductor or metallic nanomaterials, which will provide in the future a significant increase in the speed of computing systems. Extended molecules of silicon polyprismanes may be the ultrathin functional nanowires with controlled electronic characteristics and can be used in elements of computational logic. According to our studies, they have been found to be potentially good conductors. Encapsulation the metastable nanostructures (for example, nitrogen chains) inside the higher silicon polyprismanes can stabilize the latter, which will make it possible to obtain an effective high-energy material. Furthermore, high electrophilicity of polyprismanes indicates the possibility of including the charged ions inside their structure that might be useful in biological and medical applications as drug delivery systems. Also, silicon polyprismanes can be used for storing and transporting hydrogen instead of carbon nanomaterials traditionally offered in the literature for these purposes. Their conductive properties will allow one to use the polysilaprismanes as the elements of measuring equipment, for example, tips of a scanning tunneling microscope. In addition, the absence of free covalent bonds makes them insensitive to environmental pollutants (free radicals), which reduces the reactivity of the silicon needle, allowing one to achieve the atomic resolutions in the microscope.

### 3. Concluding remarks

In conclusion, we have predicted new low-dimensional $sp^3$-hybridized silicon complexes that are called polysilaprismanes. It was shown for the first time that unsubstituted silicon in the form of polysilaprismanes was able to exhibit properties qualitatively different from the properties of bulk silicon. We investigated their electronic properties and confirmed that they could be classified as metals. In addition, polysilaprismanes exhibit remarkable mechanical and chemical properties that can be extremely useful in the different areas of next-gen nanotechnologies, such as nanoelectronics, biological or medical applications, measuring equipment, energetics, etc. However, further research of specific tasks of polysilaprismane use is needed, including the determination of temperature regimes for the operation of structures based on silicon polyprismanes, their kinetic stability, toxicity, etc. We believe that our study will be of interest to the scientific community and will serve as a stimulus for further theoretical and experimental studies of unique silicon polyprismanes.

### Acknowledgments

The presented study was performed with the financial support of the Russian Science Foundation (Grant No. 18-72-00183).

**FIGURES**

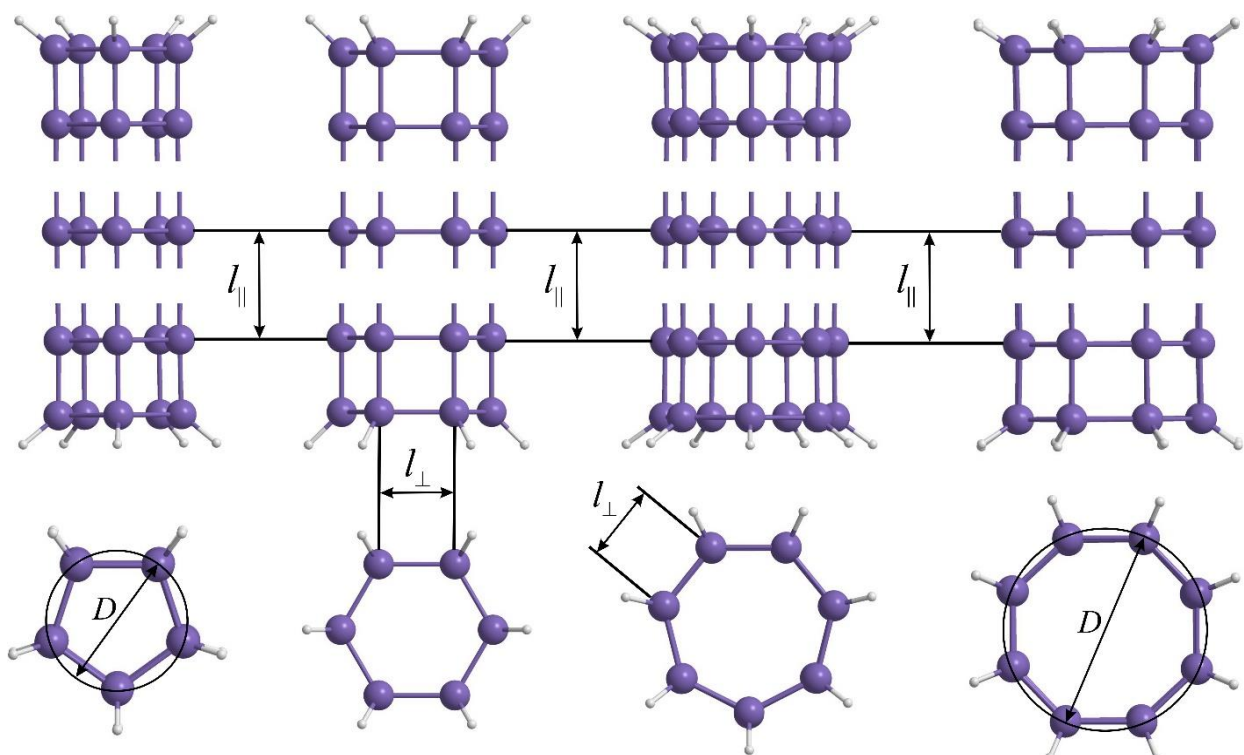

**Figure 1**. Left to right: Molecular structures of [$n$,5]-, [$n$,6]-, [$n$,7]-, and [$n$,8]silaprismanes. Top – side view, bottom – top view. Symbol $D$ corresponds to the effective diameter of silicon polyprismane, symbols $l_{\parallel}$ and $l_{\perp}$ correspond to the interlayer and intralayer Si-Si bonds, respectively.

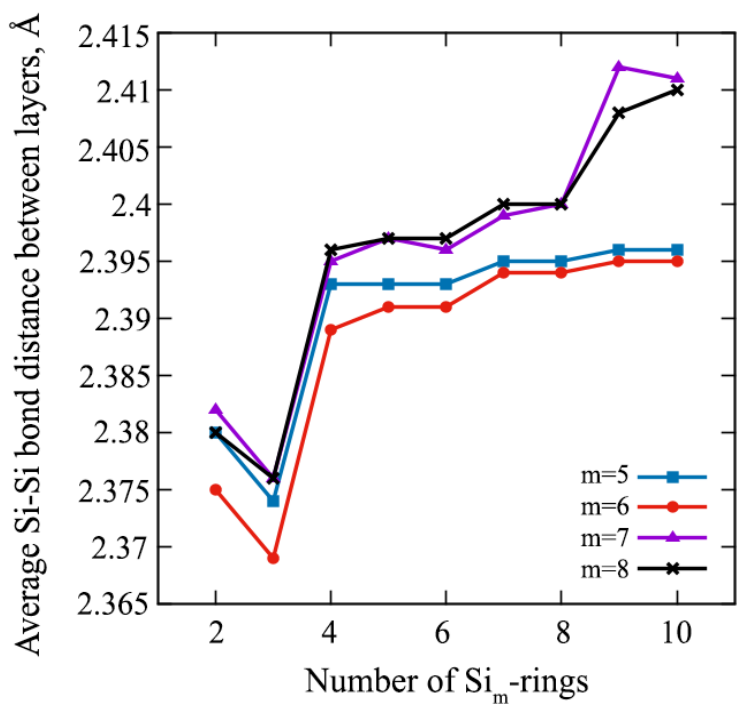

**Figure 2**. The average Si-Si bond distance between layers versus the number of silicon rings in the polyprismane obtained at the DFT/PBE0/6-311(d,p) level of theory

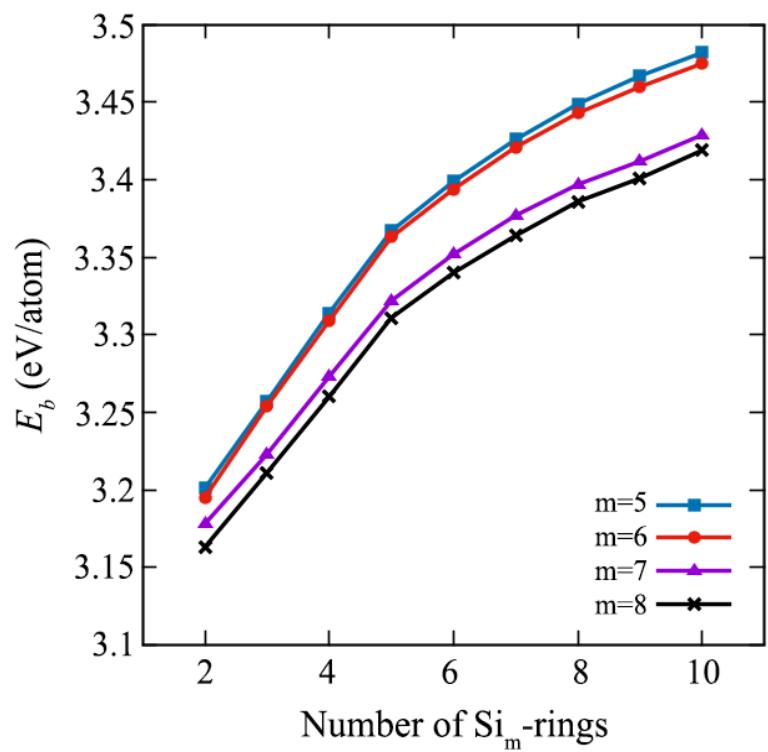

**Figure 3**. The binding energy of silicon polyprismane versus the number of silicon rings in the system obtained at the DFT/PBE0/6-311(d,p) level of theory

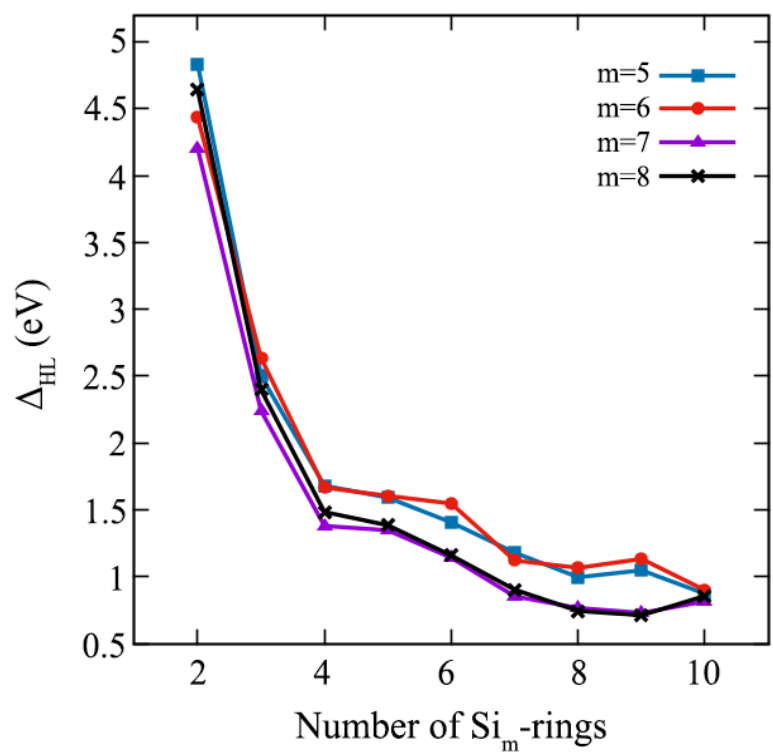

**Figure 4**. The HOMO-LUMO gap of silicon polyprismane versus the number of silicon rings in the system obtained at the DFT/PBE0/6-311(d,p) level of theory

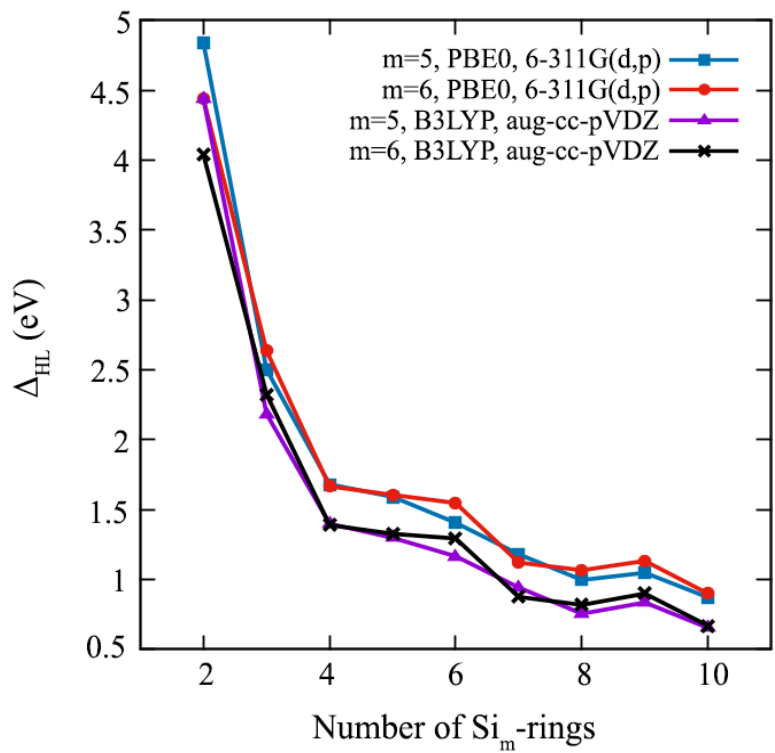

**Figure 5**. The HOMO-LUMO gap of [*n*,5]- and [*n*,6]polysilaprismanes versus the number of silicon rings in the system *n* obtained at the DFT/PBE0/6-311(d,p) and DFT/B3LYP/aug-cc-pVDZ levels of theory

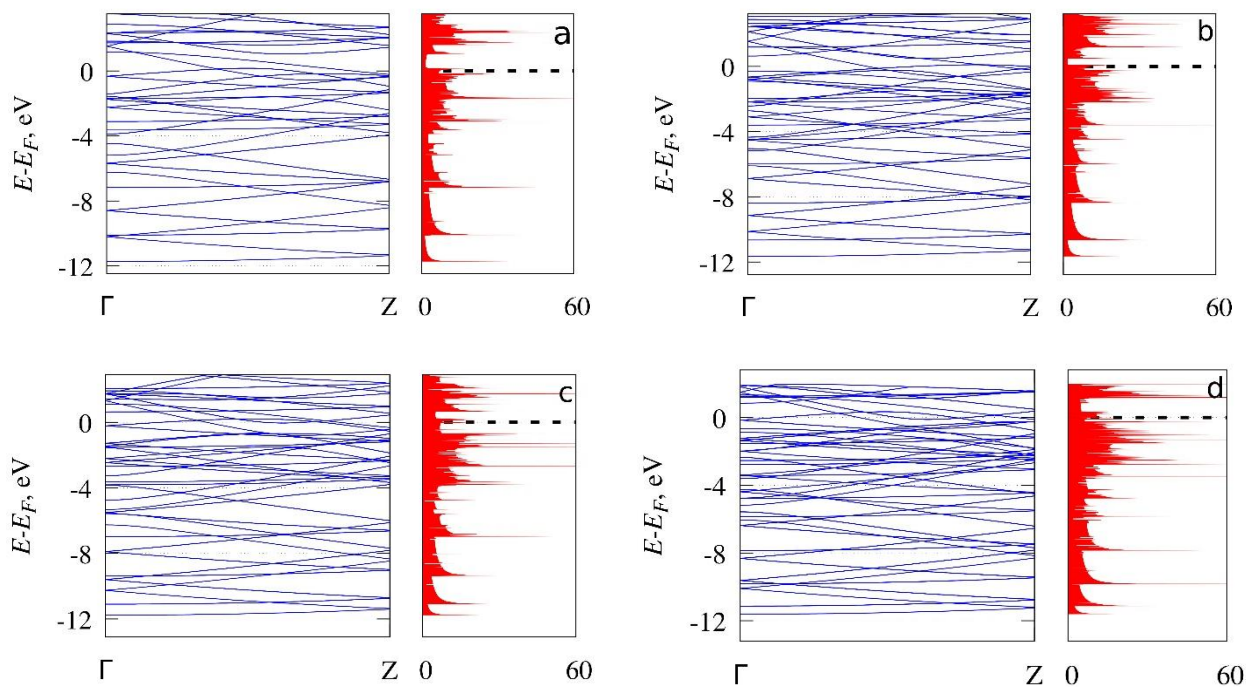

**Figure 6**. Band structures and densities of electronic states of the [*n*,5]- (a), [*n*,6]- (b), [*n*,7]- (c), and [*n*,8]- (d) polysilaprismanes

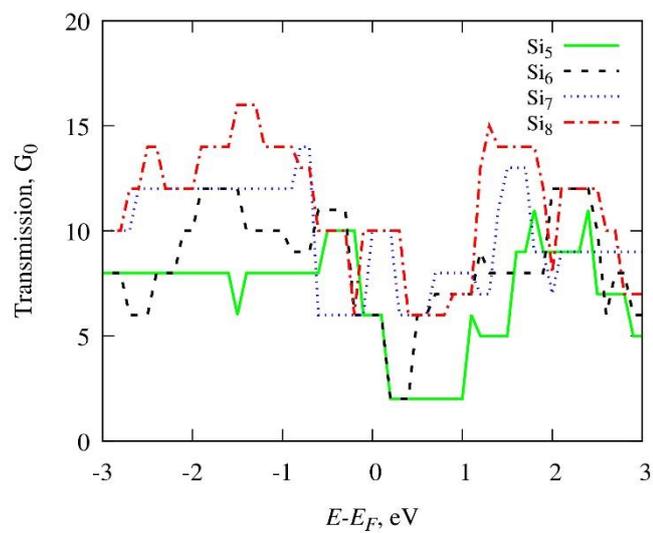

**Figure 7**. The electronic transmission coefficients of [*n*,5]- (green line), [*n*,6]- (black line), [*n*,7]- (blue line), and [*n*,8]- (red line) polysilaprismanes

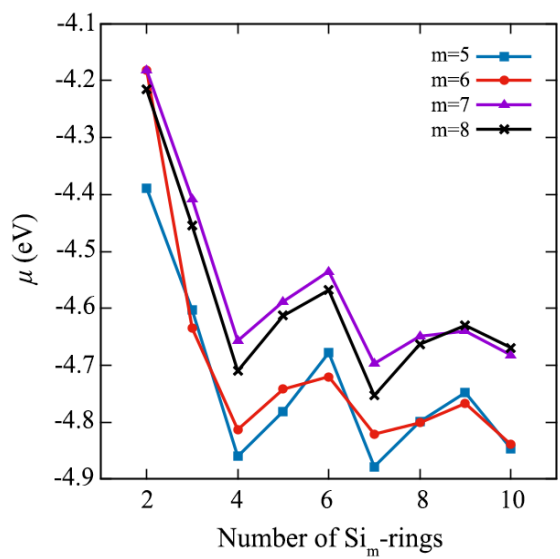

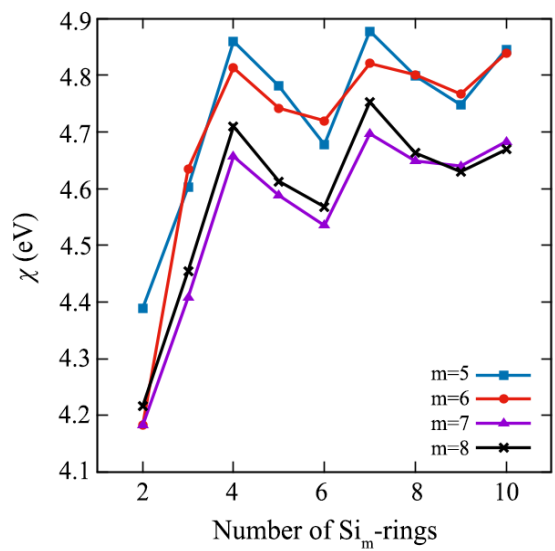

**Figure 8.** The chemical potential of silicon polyprismanes versus the number of silicon rings in the system obtained at the DFT/PBE0/6-311(d,p) level of theory.

**Figure 9.** The electronegativity of silicon polyprismanes versus the number of silicon rings in the system obtained at the DFT/PBE0/6-311(d,p) level of theory.

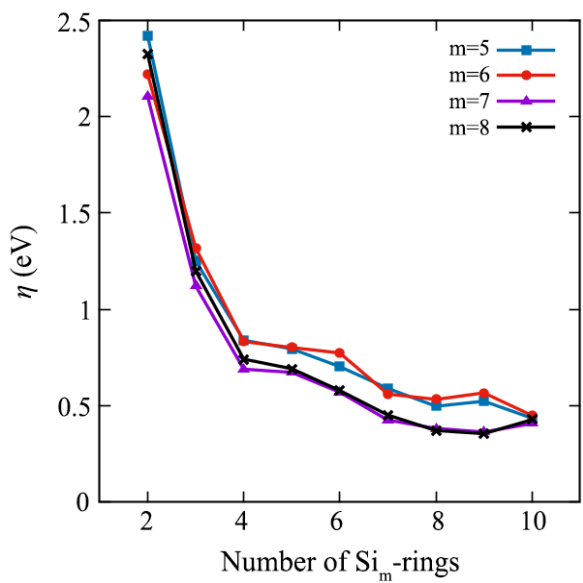

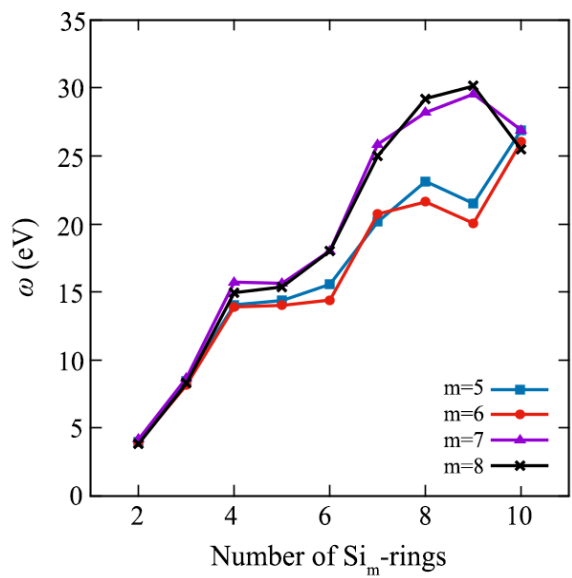

**Figure 10.** The hardness of silicon polyprismanes versus the number of silicon rings in the system obtained at the DFT/PBE0/6-311(d,p) level of theory.

**Figure 11.** The electrophilicity of silicon polyprismanes versus the number of silicon rings in the system obtained at the DFT/PBE0/6-311(d,p) level of theory.